# Sentiment-Driven Community Detection in a Network of Perfume Preferences

Kamand Kalashi, Sajjad Saed, Babak Teimourpour*
Department of Information Technology Engineering, School of Systems and Industrial Engineering, Tarbiat Modares University (TMU), Tehran, Iran
{ kalashi.kamand, saed_s, b.teimourpour}@modares.ac.ir

***Abstract*— Network analysis plays a growing role in diverse fields, including the fragrance industry, where perfume networks are formed by representing perfumes as nodes and shared user preferences as edges. Analyzing these networks through community detection can reveal clusters of similar perfumes, offering valuable insights into consumer preferences, enhancing recommendation systems, and informing targeted marketing strategies.**

**Our primary objective is to apply community detection techniques to group similar perfumes liked by users into relevant clusters, aiding in perfume recommendations. We constructed a bipartite network from user reviews on the Persian retail platform "Atrafshan," with nodes representing users and perfumes, and positive comments forming edges. This bipartite network was transformed into a Perfume Co-Preference Network, linking perfumes liked by the same users. By applying community detection algorithms, we identified clusters based on shared user preferences, contributing to a better understanding of user sentiment in the fragrance industry.**

**We enhanced sentiment analysis by incorporating emojis and a user voting system to improve accuracy. Emojis, mapped to their Persian equivalents, captured the emotional context of reviews, while user ratings for Scent, Longevity, and Sillage refined sentiment classification. Edge weights were adjusted by blending adjacency values with user ratings in a 60:40 ratio, ensuring they reflect both connection strength and user preferences. These innovations improved the modularity of the detected communities, resulting in more precise perfume groupings.**

**This research pioneers the application of community detection to perfume networks, offering new insights into consumer preferences. Our contributions in sentiment analysis and edge weight refinement provide actionable insights for optimizing product recommendations and marketing strategies in the fragrance market.**

## I. Introduction

Network analysis has experienced substantial growth across various domains, ranging from social networks to biological systems, and more recently, in commercial product networks. In the realm of perfumes, networks can be constructed with nodes representing perfumes and edges capturing relationships based on shared user preferences. By analyzing these perfume networks—where edges represent positive sentiment connections from shared user feedback — community detection can unveil groups of similar perfumes. Understanding these communities offers valuable insights into consumer preferences, enabling the improvement of perfume recommendation systems and the development of more targeted marketing strategies [1]. Such analysis helps uncover hidden patterns in user behavior, driving better business outcomes in the fragrance industry.

Community detection is a process that groups nodes in a network into clusters where the connections between nodes within each group (intra-cluster density) are denser than the connections to nodes outside the group (inter-cluster density) [2], [3] (Fig.1). In other words, community detection aims to uncover the underlying structure of networks by identifying groups of nodes (communities) that are more densely connected internally than with the rest of the network. This process is beneficial for understanding the organization of complex networks and has applications in various domains, including social networks, computer science, biological networks, marketing, and technological networks [4], [5], [6]. In recent years, community detection has found significant applications in recommender systems, where it helps to identify groups of similar items or users, thereby improving the accuracy of recommendations [2].

### A. *Importance of Community Detection in Perfumes Network*

In the context of perfume networks, community detection can reveal clusters of perfumes that are frequently reviewed or liked by similar groups of users. This information is crucial for several reasons:

- Personalized Recommendations: By identifying communities of perfumes, it becomes possible to provide more personalized recommendations to users. For example, if a user likes a perfume in a particular community, other perfumes in the same community are likely to be of interest to that user [7].

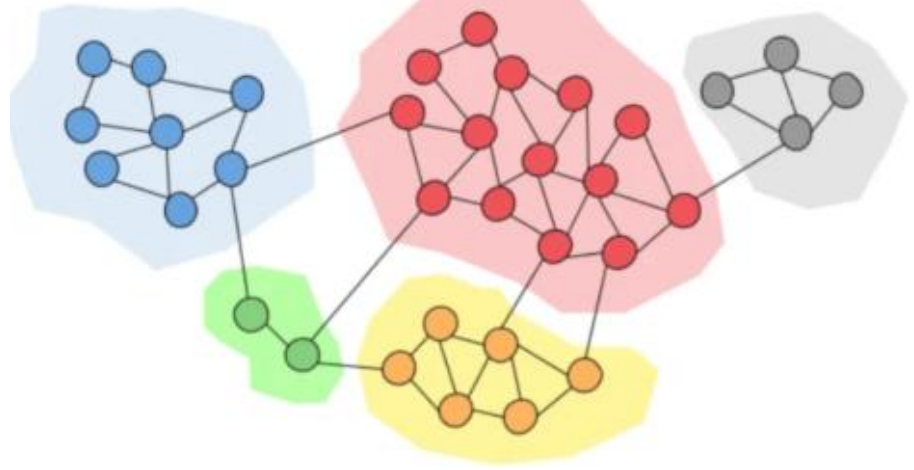

Fig. 1. A schematic representation of community detection in a network.

- Market Segmentation: Businesses can use community detection to segment the market into different user groups based on their preferences. This segmentation allows for targeted marketing strategies, where specific perfumes are promoted to the most relevant user segments [8].
- Trend Analysis: Communities in perfume networks can help identify emerging trends and popular perfumes within different user groups. This insight is valuable for manufacturers and retailers to adjust their inventory and marketing efforts accordingly [9].

### B. *Role of Sentiment Analysis*

By analyzing user comments and reviews, sentiment analysis provides insights into the emotions and opinions of users regarding different perfumes. This information can be integrated with community detection to achieve the following:

- Community Sentiment Profiling: By analyzing the sentiments expressed within each community, we can profile communities based on their overall sentiment (positive, negative, or neutral). This profiling helps in understanding the general perception of perfumes within different communities [10], [11].
- Improved Recommendations: Incorporating sentiment analysis into community detection allows for more nuanced recommendations. For instance, if a community has predominantly positive sentiment toward certain perfumes, those perfumes can be recommended more confidently to users in that community [12]-[16].
- Feedback and Improvement: Sentiment analysis provides direct feedback from users, which can be used by perfume manufacturers to improve their products. By understanding the specific aspects of perfumes that users like or dislike, manufacturers can make informed decisions to enhance product quality [17], [18].

### C. *Objectives and Contributions*

The primary objective of this study is to apply community detection techniques to group similar perfumes liked by users into relevant clusters, aiding in perfume recommendations, using a network derived from user reviews on a Persian perfume retail platform. The research begins with the collection of comprehensive data, including user comments, ratings, and key user information such as usernames and user IDs. Additional metadata was gathered, including perfume attributes (e.g., fragrance group, perfumer, and production details like brand name in Persian, country of origin, and release year). Furthermore, overall vote polls for Scent, Longevity, Sillage, and Bottle Design were recorded, along with details on whether comments were standalone or replies. This rich dataset is then utilized for further analysis.

Subsequently, sentiment analysis is performed to categorize the comments as positive or negative. This analysis serves as the basis for constructing a bipartite network, with users and perfumes as nodes and positive comments as edges connecting users to perfumes. To specifically focus on perfume relationships, the bipartite network is transformed into a Perfume Co-Preference Network, where perfumes as nodes are linked if they receive positive feedback from the same users.

Community detection algorithms are then applied to this Perfume Co-Preference Network, facilitating the identification of clusters of similar perfumes based on user preferences. These clusters offer valuable insights into user sentiment and preferences, which can be utilized to enhance recommendation systems and inform business strategies within the fragrance industry.

Thus, our contributions can be summarized as follows:

- **First-Time Application of Community Detection in Fragrance Analysis**: This research marks the first application of community detection methods to perfume networks, establishing a novel approach to analyzing consumer preferences within the fragrance industry. By pioneering this research, we provide valuable insights into how user sentiment and preferences can be effectively represented and understood through network analysis, filling a significant gap in the existing literature on perfume studies.

- **Dataset Extraction**: For the first time, this study extracts and makes available a rich dataset from a Persian perfume retail platform, including user reviews, ratings, and perfume attributes. This dataset is hosted on GitHub and is made available for further analysis, supporting deeper research into consumer preferences and behaviors in the perfume industry.

- **Innovations in Sentiment Analysis**: We introduced several innovations in sentiment analysis, particularly by incorporating emojis and a user rating system to enhance the accuracy and depth of our results.

  Emojis, commonly used in Persian user reviews to convey emotions, were mapped to their corresponding Persian phrases, enriching the analysis by capturing the full emotional context of comments. For example, emojis like "😍" were translated into phrases like "چشم های عاشق" ("loving eyes"), ensuring that the sentiment conveyed by visual elements was included in the text analysis.

  Additionally, we integrated a voting system that leveraged user ratings for perfume attributes such as Scent, Longevity and Sillage. These ratings refined the sentiment classification results by assigning positive or negative labels based on the scores, where high ratings (7 or above) indicated positive feedback and low ratings (3 or below) indicated negative feedback. This combination of textual sentiment and voting-based refinement allowed for a more nuanced analysis of user preferences.

Ultimately, we used this enhanced sentiment analysis to construct three Perfume Co-Preference Networks, each biased towards one of the user rating categories based on the sentiment analysis results.

- **Refinement of Edge Weights**: A notable contribution of this study is the integration of overall vote polls to refine the edge weights of the Perfume Co-Preference Network. Average total user ratings across essential categories—Scent, Longevity, and Sillage—were collected to quantify preferences associated with each perfume. For instance, aggregate ratings were recorded, such as an average Scent rating of 8.2 based on 285 votes.

  To assign edge weights, the original adjacency matrix values, representing the number of users who commented positively on both perfumes, were combined with the weights from the relevant category being analyzed—such as Sillage, Scent, or Longevity—reflecting overall user satisfaction for that specific attribute. The edge weight connecting two perfumes was determined by prioritizing the category-specific weights over the original adjacency values, using a ratio of 60% for the category-specific weights and 40% for the original adjacency values. This approach ensures that the final weight represents a blend of the original connection strength and the selected category's scores of the connected perfumes, thereby adequately representing user sentiment in the network.

  Following this, community detection algorithms were applied to the Perfume Co-Preference Networks, facilitating the identification of clusters of similar perfumes based on shared user preferences. This innovative approach enhances the accuracy and meaningfulness of the clusters generated by the community detection algorithms, ultimately improving insights into consumer preferences and informing strategies within the fragrance industry.

- **Impact on Modularity**: The inclusion of sentiment classification, where we construct the network based solely on positive comments, combined with the refinement of edge weights, contributed to a significant increase in the modularity score across all of our community detection techniques compared to the primary network. These enhancements allowed for clearer and more meaningful partitions within the network, resulting in stronger associations of perfumes within the identified clusters based on shared user preferences. This approach not only improves the precision of perfume recommendations but also provides deeper insights into consumer preferences and behaviors in the fragrance industry.

The paper is organized to enhance comprehension and facilitate navigation through its content. Section 2 provides a review of related works, while Section 3 outlines the methodology and proposed model. Section 4 presents a discussion and analysis of the results. Section 5 concludes with a summary of the key findings and recommendations for future research directions. Finally, Section 6 acknowledges individuals who contributed to the research.

## II. Related work

### A. Community Detection in Social Networks

Community detection in social networks has been extensively studied due to its importance in understanding the structure and dynamics of social interactions. Fortunato [5],[19] provides a comprehensive review of community detection methods, highlighting various algorithms and their applications. In this paper, we will review four major types of community detection methods in the related work section, including Divisive algorithms, Modularity-based methods, Dynamic algorithms, and the Spectral Clustering method.

- **Divisive Algorithms** take an intuitive approach to community detection by identifying and removing edges that connect different communities, progressively disconnecting the network into smaller clusters. The key challenge is identifying which edges are likely to bridge distinct communities. This approach mirrors traditional hierarchical clustering, but instead of grouping similar nodes, it focuses on removing inter-cluster edges [5].

  One of the most well-known divisive methods is the Girvan-Newman algorithm [4],[20], which relies on edge betweenness centrality. Edge betweenness measures how often an edge lies on the shortest paths between pairs of nodes, with edges that frequently connect different parts of the network having high betweenness scores. By progressively removing edges with the highest betweenness, the algorithm splits the network into distinct communities. The result is typically represented by a dendrogram, a tree diagram illustrating the successive splits. Although effective, the Girvan-Newman method is computationally intensive, with a time complexity of $O(E^2N)$, where $N$ is the number of nodes and $E$ is the number of edges, making it less practical for large-scale networks, such as the one in our research [4],[21].

- **Modularity-based methods** focus on optimizing the modularity score, which measures the density of edges within communities compared to a null model that assumes random connections [19]. The most well-known of these methods is based on maximizing the modularity function, introduced by Newman and Girvan [20]. Modularity $Q$ is calculated as:

$$Q = \frac{1}{2m}\sum_{ij}\left(A_{ij} - \frac{k_i k_j}{2m}\right)\delta(C_i, C_j)$$

  where $A_{ij}$ is the adjacency matrix, $k_i$ and $k_j$ are the degrees of nodes $i$ and $j$, and $m$ is the number of edges.

The term $\delta(C_i, C_j)$ is the Kronecker delta, which is a function used to check if two nodes i and j belong to the same community. It is defined as:

$$\delta(C_i, C_j) = \begin{cases} 1 & \text{if } i \text{ and } j \text{ are in the same community} \\ 0 & \text{otherwise} \end{cases}$$

In other words, $\delta(C_i, C_j)$ is equal to 1 when both nodes $i$ and $j$ belong to the same community $C$, and 0 when they belong to different communities. The goal is to maximize $Q$, which reflects how much better the partitioning is compared to a random network. Two prominent modularity-based methods are the Louvain and Fastgreedy algorithms.

The Louvain method (also referred to as the Multilevel method), introduced by Blondel et al. [22], employs a greedy, hierarchical approach to modularity optimization. Initially, each node is assigned to its own community. Nodes are iteratively moved to the communities of their neighbors to maximize modularity. Once no further improvement can be made, the communities are merged into "super-nodes," and the process is repeated. This multilevel approach continues until no additional increase in modularity is possible. The Louvain method is highly scalable and efficient for large networks, with a time complexity of $O(N\log N)$ [21],[23].

The Fastgreedy algorithm, proposed by Clauset et al. [24], also optimizes modularity through a greedy approach. It begins by treating each node as an individual community and evaluates the potential modularity improvement for each pair of communities. The pair of communities that provides the maximum modularity increase is merged. This process is repeated until no further improvement is possible. The Fastgreedy algorithm is particularly effective for sparse, hierarchical networks and has a time complexity of $O(N\log^2(N))$ [21],[24].

- **Dynamic Algorithms** for community detection use the concept of running dynamical processes on networks, such as diffusion, random walks, and spin dynamics, to identify community structures. The underlying idea is that the behavior of these dynamic processes will naturally reveal the boundaries of communities due to differences in interaction patterns within and between communities. Random walk dynamics are a common approach. In a network with well-defined community structures, a random walker is likely to stay within the same community for a long time before finding an edge that connects to another community. This behavior allows the detection of clusters by observing how long random walkers remain in specific groups of nodes [5]. Two prominent dynamic methods for community detection are Infomap and Walktrap.

Infomap Proposed by Rosvall et al. [25,26], identifies communities by analyzing how information flows through a network using random walks [27]. The method models the network as a communication system where the goal is to encode the network into modules (communities) in a way that maximizes information transfer efficiency. The smaller the number of encoded modules needed to describe the network, the more accurately the community structure is captured. Infomap aims to minimize the description length of a random walk, which includes two components: one for movements within communities and another for movements between them. The shorter the overall description, the better the community structure is represented. This method can be applied to both weighted and unweighted networks and has a time complexity of $O(E)$[21],[28].

Walktrap, introduced by Pons and Latapy [29], is based on the idea that short random walks tend to remain within the same community. Walktrap starts with an initial partition where each node is its own community. Then, it calculates the "distance" between nodes based on random walks, merges adjacent communities that are close to each other, and updates the distances. This process is repeated until only one community remains, and the resulting hierarchical structure can be used to find the optimal partitioning. The computational complexity of Walktrap is $O(N^2\log(N))$ for sparse networks [21],[23].

Another dynamic algorithm often used for community detection is based on spin dynamics, such as the Spinglass algorithm [30]. This method models the network as a system of interacting spins, where nodes are assigned spin states, and the goal is to minimize a Potts model Hamiltonian [31]. Nodes with similar spin states are grouped into the same community, while nodes with different states are placed in different communities. Simulated annealing [32] is typically used to minimize the energy of the system [33], resulting in community structures that correspond to the ground state of the spin glass model. The computational complexity of this approach is approximately $O(N^{3.2})$ for sparse graphs [21],[34].

In this algorithm, an important parameter is gamma (γ), which controls the resolution of the detected communities. By adjusting gamma, the algorithm can identify either larger, broader communities (with a lower gamma value) or smaller, more granular ones (with a higher gamma). Essentially, gamma dictates the balance between intra-community connections and inter-community connections. A low gamma allows for more intra-community edges, potentially merging smaller communities into larger ones, while a high gamma forces the algorithm to favor smaller, tighter communities.

Dynamic methods, particularly those based on random walks, are advantageous because they consider the flow of information through the network rather than relying purely on structural properties like edge density. This allows them to capture more nuanced community structures, especially in directed or weighted networks, where the dynamics of node interactions play a significant role in community formation. Moreover, these algorithms are useful in detecting overlapping communities, as random walkers may traverse multiple clusters [19].

- **Spectral clustering method** is a technique for partitioning objects into clusters using eigenvectors of matrices associated with the data [5]. Given $n$ objects $x_1, x_2, \ldots, x_n$ and a symmetric, non-negative similarity function $S$ (where $S(x_i, x_j) = S(x_j, x_i) \geq 0$ ), spectral clustering transforms the data based on the eigenvectors of the similarity or Laplacian matrix, enhancing cluster visibility that might not be evident in the original space [35],[36]. The clustering process involves several key steps:
  1. Compute the Laplacian matrix $L$.
  2. Calculate the eigenvectors corresponding to the smallest $k$ eigenvalues of $L$.
  3. Form the $n \times k$ matrix $V$ from these eigenvectors.
  4. Use $V$ to represent graph vertices in a $k$-dimensional space.
  5. Apply $k$-means clustering to identify clusters.

  This approach is advantageous for identifying non-convex clusters, which traditional methods like $k$ means may miss [37].

  Spectral clustering can be executed using unnormalized or normalized Laplacians. The unnormalized approach, proposed by Shi and Malik [35], utilizes the Laplacian derived from the adjacency matrix, while normalized methods account for vertex degrees, enhancing robustness [38]. For instance, Shi and Malik's random walk Laplacian $L_{rw} = I - D^{-1}A$ stabilizes eigenvector representations in graphs with low-degree vertices.

In the context of social recommender systems, community detection helps in identifying groups of users with similar preferences, thereby improving recommendation accuracy. Gasparetti et al. [2] surveyed various techniques for community detection in social recommender systems, emphasizing the importance of understanding user interactions and preferences. Yang et al. [39] applied Bayesian models to detect communities in social networks, which enhanced the recommendation performance by capturing latent user preferences.

### *B. Perfume Networks and Recommender Systems*

The study of perfume networks is relatively new, with limited research focusing on this specific domain. These networks provide valuable insights into consumer preferences and market trends. Despite the scarcity of literature specifically on perfume networks, related studies on product networks and user-generated content can provide useful insights. One relevant study by Gao [40] examined product co-purchasing networks on e-commerce platforms, revealing the community structures of products frequently bought together. This approach can be adapted to perfume networks by analyzing user reviews and preferences. Similarly, studies on sentiment analysis in product reviews, such as the work by Pang and Lee [12], provide methods for extracting sentiments from user comments, which can be integrated with community detection to enhance understanding of user preferences.

### *C. Sentiment Analysis*

Sentiment analysis is a crucial tool for understanding user opinions and emotions expressed in text data. In the context of perfume networks, it reveals overall sentiment toward different perfumes, providing valuable feedback for manufacturers and marketers. Foundational work by Pang and Lee [12] outlines techniques for classifying sentiments in text, while Cambria [18] advances the field with methods for affective computing and sentiment analysis across various applications. Recent studies have integrated sentiment analysis with community detection to enhance the understanding of user communities. For example, Wang et al. [10] demonstrate how sentiment analysis can profile communities based on their overall sentiment, improving recommendation accuracy. This integration is particularly relevant for perfume networks, where user comments often contain rich information about personal preferences and emotions.

In our paper, we apply sentiment analysis to categorize user comments from a Persian perfume retail website into positive and negative sentiments. This classification forms the basis for constructing a bipartite network, linking users and perfumes through positive feedback. We then transform this network into a Perfume Co-Preference Network, connecting perfumes that receive positive reviews from the same users. By applying community detection algorithms to this network, we identify clusters of similar perfumes, offering insights into user preferences.

We leverage ParsBERT [41], a state-of-the-art pre-trained transformer model for Persian language understanding, to perform sentiment analysis on user reviews. ParsBERT has shown superior performance in Persian NLP tasks, including sentiment analysis, text classification and named entity recognition, outperforming multilingual models like BERT in accuracy and efficiency.

Its architecture, utilizing Bidirectional Encoder Representation from Transformers (BERT), is optimized for Persian, accommodating unique linguistic properties such as complex morphology and distinct syntactic structures [42].

This advantage is crucial as multilingual models, including Multilingual BERT, often struggle with non-Latin languages like Persian, as their shared representations fail to capture language-specific nuances. ParsBERT addresses this challenge by being trained on a large corpus of Persian texts, providing accurate context-based representations essential for sentiment analysis, where subtle language differences significantly impact outcomes [41],[42].

Overall, by using a model specifically tailored to the Persian language, we push the boundaries of Persian NLP tasks, delivering a highly personalized experience in perfume recommendations, traditionally dominated by English-centric models.

## III. Methodology

### A. Data Collection

The Data collection was conducted using the Atrafshan[1] website, a popular Iranian platform for perfume Sales, reviews and ratings. A total of 36,434 comments were extracted from 7,387 unique users and 1,369 unique perfumes. The objective was to extract user comments, ratings, and relevant information about the most viewed perfumes (Fig.2).

Specific perfume pages (Fig.3) were targeted by analyzing the webpage structure to locate the comment section and determine the total number of comments available for each fragrance. Based on this, the number of pages to be scraped for each perfume was calculated, enabling systematic navigation through multiple pages of user feedback. Key user information, including usernames, user IDs, comments, and the user's voting on four categories—Scent, Longevity, Sillage (the scent trail a perfume leaves behind as it evaporates), and Bottle Design—was meticulously recorded while avoiding duplicate entries. For instance, a typical user comment (Fig.4) would include ratings such as "10 for Scent, 10 for Longevity, 8 for Sillage, and 9 for Bottle Design." In addition, perfume names and perfume IDs were collected to uniquely identify each fragrance in the dataset, ensuring accurate linking between user comments and perfumes.

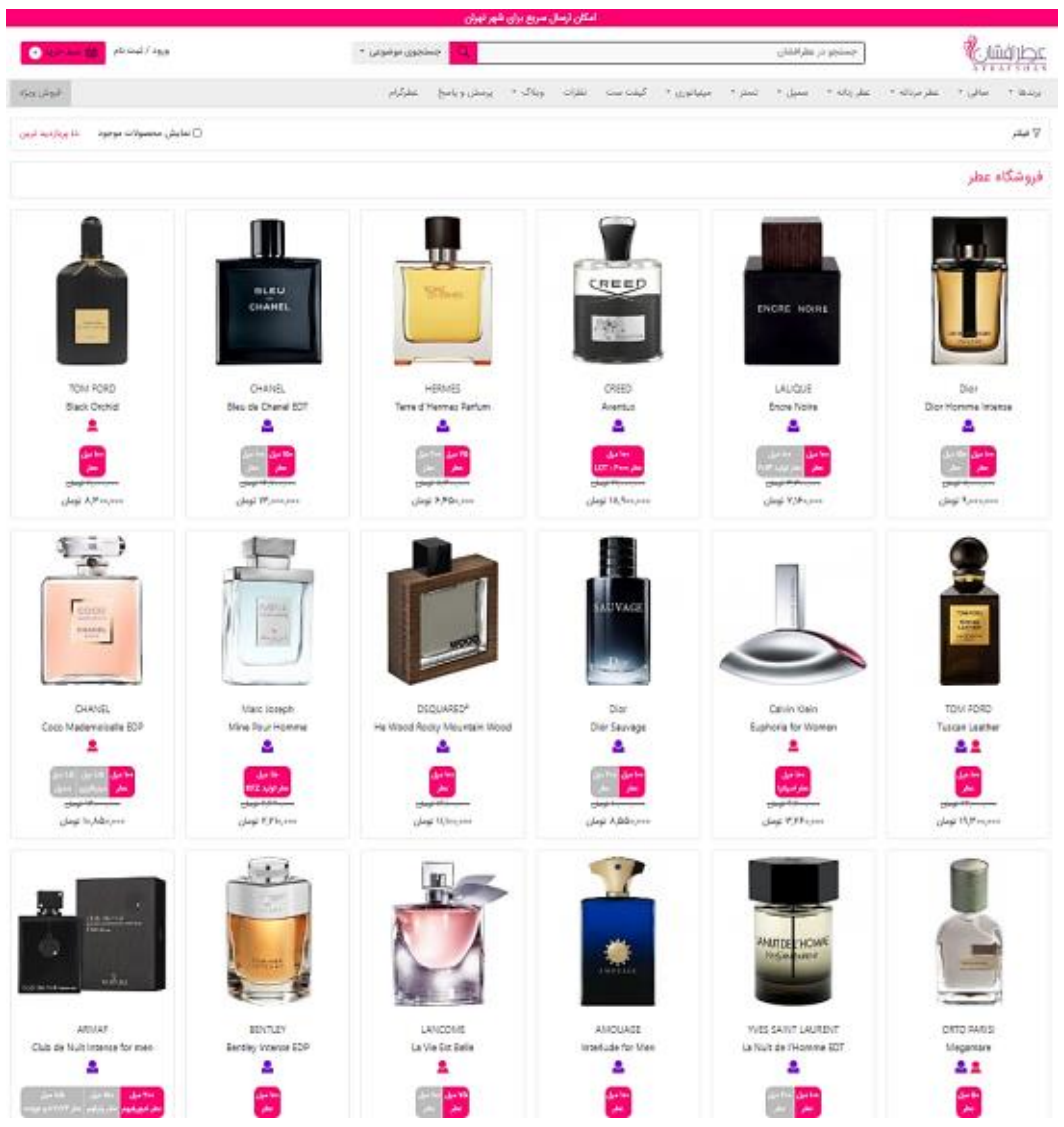

Fig. 2. First page of Most Viwed Perfumes of Atrafshan Website.

Detailed information about each perfume was collected, such as fragrance specifications (including fragrance group, perfumer, and nature), and production details (such as the brand name in Persian, country of origin, and release year). For example, a perfume might belong to the "Woody Floral Musk" group, with *François Demachy* as the perfumer, and have a warm nature. A direct URL to the perfume's page on Atrafshan was also captured.

Each perfume's page also featured overall vote polls, including the total number of users who participated in voting for categories like Scent, Longevity, Sillage, and Bottle Design. For instance, for the *Lancome La Vie Est Belle* perfume, the overall voting was as follows (Fig.5):

-Scent: 8.1 (based on 96 votes)
-Longevity: 8.5 (based on 139 votes)
-Sillage: 8.6 (based on 135 votes)
-Bottle Design: 8.3 (based on 96 votes).

It was also recorded whether a user comment was a reply to another comment or a standalone review. The collected data was periodically saved locally to maintain its integrity and prevent loss, resulting in a structured dataset for further analysis.

The scraping process was designed with a focus on error handling and resilience. Several mechanisms were implemented to ensure uninterrupted data collection despite potential issues with individual pages. Anticipating broken links, missing data, or server timeouts, the script included retry functions to reload problematic pages after a brief interval. Pages that continued to encounter errors were logged and bypassed, allowing the process to proceed without disruption. The script also utilized `time.sleep()` method to introduce delays between requests, preventing server overload and minimizing the risk of being blocked for excessive querying. This approach ensured that each page was accessed thoughtfully, reducing strain on the server.

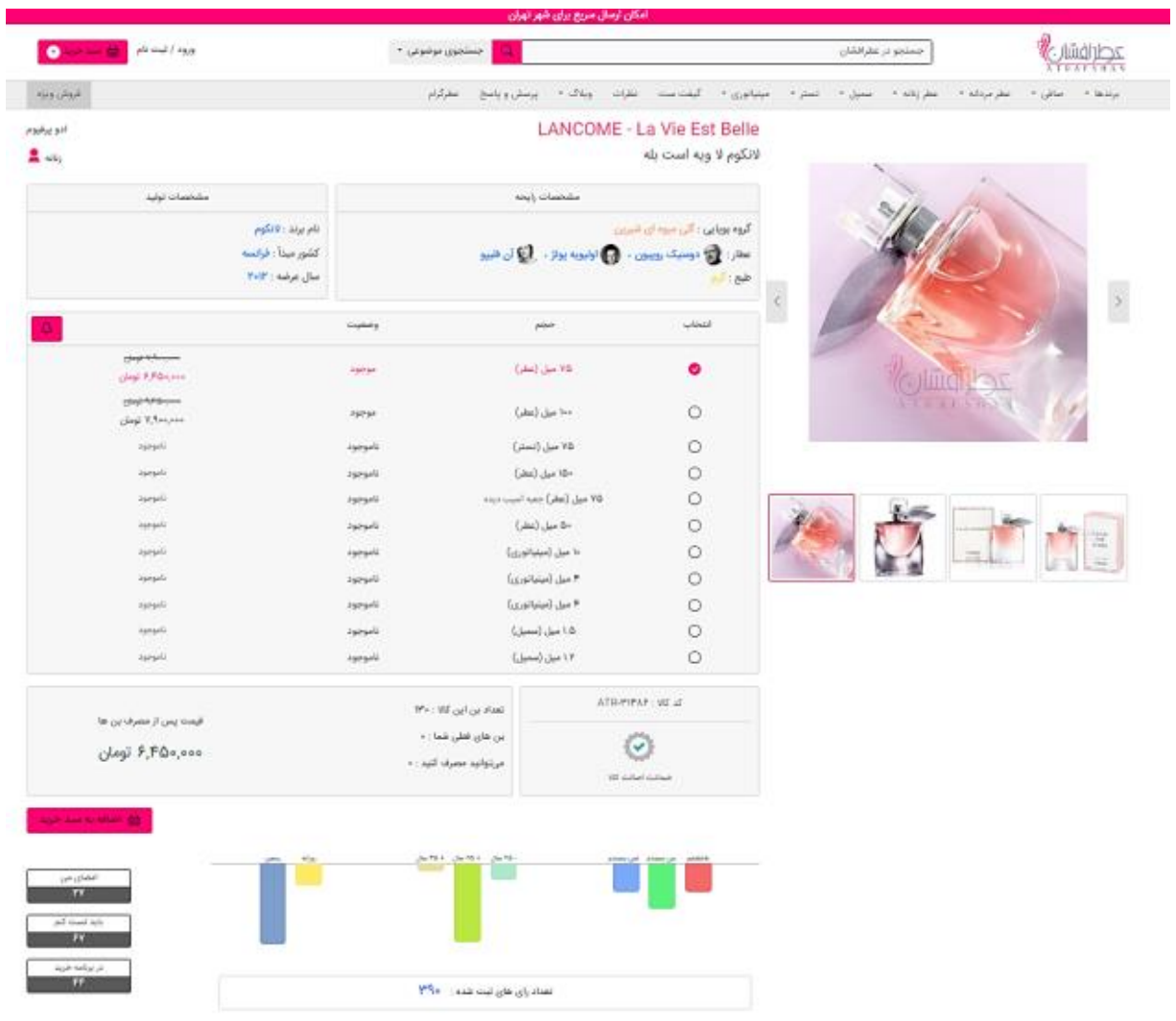

Fig. 3. *Lancome La Vie Est Belle* perfume page on Atrafshan.

[1] https://atrafshan.com/

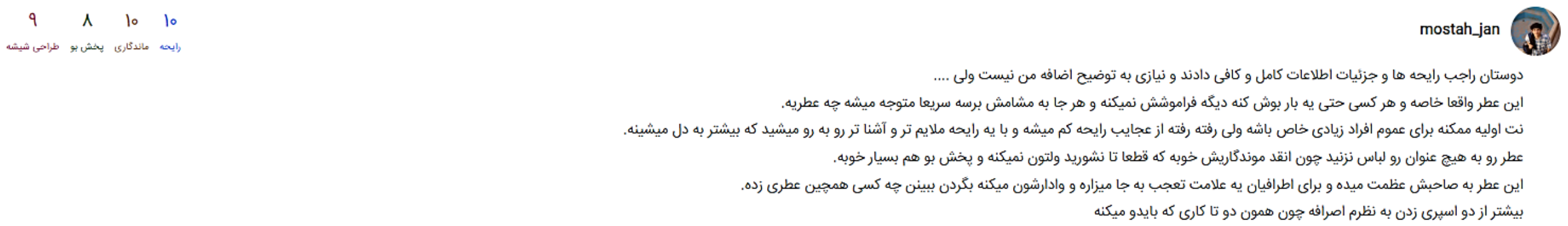

Fig. 4. A Typical User Comment that includes Votings: (From Right to Left)"10 for Scent, 10 for Longevity, 8 for Sillage, and 9 for Bottle Design."

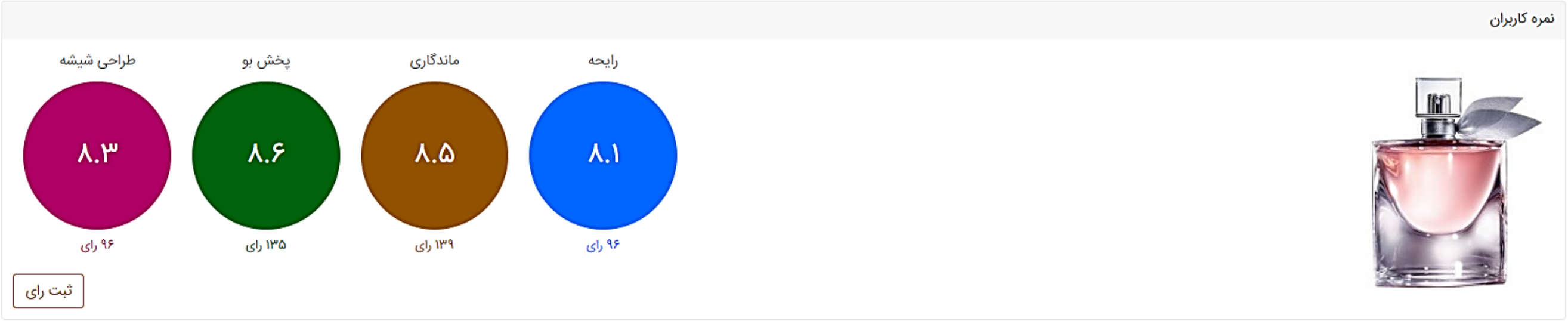

Fig. 5. Average total user ratings for *Lancome La Vie Est Belle* perfume: (From Right to Left) "8.1 for Scent, 8.5 for Longevity, 8.6 for Sillage, and 8.3 for Bottle Design."

Furthermore, missing or incomplete data, such as absent user comments or ratings, were managed by the script to skip these gaps without causing failures. Regular backups of the data were performed throughout the process, safeguarding against unexpected interruptions. These precautions, combined with robust error-handling features, resulted in a seamless and efficient data collection operation, capturing a wide array of user feedback from perfume reviews.
Additionally, we have shared our dataset with the community in our GitHub repository[2], providing access to the collected data for further research and analysis.

### B. *Data Preprocessing*

The preprocessing of data involved several key steps to ensure the integrity and reliability of the user-perfume network used for analysis. This section outlines three main preprocessing techniques applied to the dataset:

- **Unique User and Perfume IDs:** To prevent any overlap or confusion between user IDs and perfume IDs, unique prefixes were assigned to each. User IDs were prefixed with "user_" and perfume IDs with "perfume_", ensuring they were treated as distinct entities in the network. This approach eliminated the possibility of mistaking user IDs for perfume IDs, a potential issue when numerical identifiers overlap between different entities. For example, the user ID “4225” would become “user_4225” and the perfume ID “4225” would become “perfume_4225” ensuring clear differentiation. This process also included a validation step to check for any overlap between prefixed user IDs and perfume IDs, confirming no duplicates existed.

- **Handling Missing and Duplicate Values:** The dataset was carefully examined for missing or NaN values, particularly focusing on user and perfume information. Rows with missing User ID or Perfume ID were removed to maintain the accuracy of the dataset. This helped ensure that gaps in the data did not compromise the integrity of the analysis. Additionally, duplicate user-perfume interactions were identified and eliminated to avoid redundant entries.

- **Filtering Perfumes with Low User Engagement:** To enhance the reliability of the bipartite user-perfume network, perfumes that received three or fewer total comments, regardless of sentiment, were excluded from the analysis. This criterion ensured that only perfumes with a significant level of user engagement were included, thereby improving the robustness of the network. Perfumes with fewer comments might not provide enough data to accurately reflect user preferences and could introduce noise into the analysis. By filtering out these perfumes, we focused on well-regarded products and laid the foundation for building a reliable Perfume Co-Preference Network. This careful curation of connections allowed for more meaningful and accurate community detection and clustering. In Table I., the statistics of the primary and filtered user-perfume network data are provided.

[2] https://github.com/Kalashi-Saed-Collaborations/SentimentDrivenCommunityDetection

### *C. Sentiment Analysis*

In this section, we detail the sentiment analysis performed on user comments regarding perfumes, focusing on their emotional tone and how it relates to key aspects of the perfumes. The analysis classified user comments into either positive or negative categories, reflecting the overall sentiment expressed by the users. This classification was carried out using a combination of text preprocessing, sentiment classification with a pre-trained ParsBERT model, and user voting integration to enhance classification accuracy.

- **Text Preprocessing:** The preprocessing phase is a critical step in ensuring that the sentiment analysis captures the nuances of user comments effectively. This phase included several important steps:
  - **Normalization with Hazm[3]**: To standardize the Persian text, we employed a dedicated preprocessing tool known as Hazm. This library is specifically designed for Persian language processing and provides a range of functionalities that address common challenges in text normalization. This process rectified various issues, including misspellings related to blank space. By addressing such discrepancies, Hazm ensures that the text is consistent, making it easier to analyze. The normalization process also includes converting numerals to a standard format, normalizing diacritics, and ensuring consistent use of punctuation marks, which collectively enhance the accuracy of subsequent analyses.
  - **Emoji Mapping**: To account for the emotional nuances conveyed by emojis in user comments, we constructed a dictionary that maps common emojis to their respective Persian phrases. Noticing that a significant number of comments in our dataset used emojis, we recognized the importance of translating them. In total, 392 emojis were translated, allowing for a richer understanding of the sentiment behind comments that use emojis, leading to a better comprehension of the text and improving our sentiment classification results with fewer errors. For instance, an emoji like "😍" could be translated to "چشم های عاشق" (meaning "loving eyes"), thereby preserving the emotional context.To ensure the accuracy of our understanding of emojis, we utilized Emojipedia[4], an emoji reference website that documents meanings and common usages of emoji characters. This resource helped us verify our interpretations before translating them into Persian. Additionally, we have shared our emoji-to-Persian dictionary with the community in our GitHub repository, which is linked in the Data Collection section of our paper. In Table II., we provide examples of common emojis along with their corresponding Persian phrases and English translation.
  - **Emoji Replacement**: A dedicated function was implemented to identify and replace emojis in the comments with their corresponding Persian phrases. The function not only replaces the emojis but also appends these phrases to the end of the comments. This ensures that the emotional sentiment associated with the emojis is retained and integrated into the textual data. By appending the Persian phrases with "و" (meaning "and") between them, we can maintain the context and flow of the original comment.

- **Sentiment Classification**: We leveraged the ParsBERT model[5], a variant of the BERT architecture that has been specifically fine-tuned for Persian text processing. The methodology for sentiment classification included the following key steps:
  - **Loading the Model**: We initialized the pre-trained ParsBERT model and its tokenizer from the Hugging Face Transformers library. This model is well-suited for understanding the intricacies of Persian language semantics and sentiment.
  - **Creating a Sentiment Classifier**: We designed a feedforward neural network to classify sentiments into binary categories: Positive and Negative. The model takes the processed comments as input and predicts their sentiment based on learned representations of the text. The classifier is trained to differentiate between the emotional undertones of various comments effectively.
  - **Batch Processing**: To manage the analysis of large datasets efficiently, we processed user comments in batches. This approach allows for optimal memory usage and speeds up the overall classification process. Each batch undergoes preprocessing, including emoji replacement and text normalization, before being fed into the sentiment classifier.

TABLE I. STATISTICS OF THE PRIMARY AND FILTERED BIPARTITE USER-PERFUME NETWORK DATA

| | ***Comments*** | ***Unique Users*** | ***Unique perfumes*** |
|---|---|---|---|
| *Primary dataset* | 36,434 | 7,387 | 1,369 |
| *Removing perfumes with 3 or fewer comments* | 35,910 | 7,387 | 1,104 |

[3] https://github.com/roshan-research/hazm

[4] https://emojipedia.org/

[5] https://github.com/hooshvare/parsbert

- **Integration of User Voting:** To further refine the sentiment classification $S$ (the sentiment label assigned to a perfume aspect), we incorporated user votes regarding three essential categories that influence the overall perfume experience: Scent, Longevity, and Sillage. This integration enhances the accuracy of sentiment labels based on direct user feedback. The process involved:
    - **Updating Sentiments**: For each user vote, we focused on the ratings $R$ (the user rating assigned to a specific aspect of the perfume) in each of the three categories, Scent, Longevity and Sillage. This approach allowed us to create three distinct CSV files of sentiment analysis, each biased towards one of the categories. The decision rules applied for updating sentiment classifications were consistent across all categories and can be summarized as follows:

$$S(R) = \begin{cases} \text{Positive} & \text{if } R \geq 7 \\ \text{Negative} & \text{if } R \leq 3 \\ \text{Ambiguous} & \text{if } 4 < R < 7 \end{cases}$$

In this schematic:

-A rating $R$ of 7 or higher results in a Positive sentiment classification $S$, indicating a strong preference for that aspect of the perfume.

-A rating $R$ of 3 or lower yields a Negative sentiment classification $S$, reflecting dissatisfaction with that aspect.

-Ratings falling between 4 and 6 are classified as Ambiguous. In this case, no change is made to the sentiment classification. This ensures that ratings deemed neutral or unclear do not dilute the reliability of sentiment analysis, allowing for more precise interpretations of consumer preferences.

This systematic integration of user voting across all three categories not only enhances the reliability of sentiment classification but also allows for a more nuanced understanding of consumer preferences across different dimensions of the perfume experience. The resulting datasets provide valuable insights into how users perceive each aspect of perfumes, enabling a more informed recommendation system.

Table III. reports the statistics of the sentiment analysis classification results, and Fig. 6 and Fig. 7 provide two examples of comments that were initially labeled as Positive and Negative, respectively, which then changed classification to Negative and Positive due to the integration of user votes. We also share the results of our sentiment analysis for each of the categories with the community in our GitHub repository, which is linked in the Data Collection section of our paper.

TABLE II. EXAMPLE EMOJI MAPPING: SELECTED EMOJIS WITH CORRESPONDING PERSIAN PHRASES AND ENGLISH TRANSLATIONS (FULL EMOJI-TO-PERSIAN DICTIONARY IS AVAILABLE ON OUR GITHUB REPOSITROY, LINKED IN THE DATA COLLECTION SECTION OF OUR PAPER.)

| *EMOJI* | *CORRESPONDING PERSIAN PHRASE* | *ENGLISH TRANSLATION* |
|---|---|---|
| 👍 | پسندیدن | LIKE |
| 😡 | عصبانی | ANGRY |
| ❤️ | عشق | LOVE |
| 👌 | عالی | GREAT |
| 😞 | ناامید | DISAPPOINTED |
| 🤢 | حالت تهوع | NAUSEATED |

TABLE III. STATISTICS OF SENTIMENT CLASSIFICATION RESULTS

| | | *Positive* | *Negative* |
|---|---|---|---|
| *ParsBERT* | | 28,757 | 7,674 |
| *Integration of User Voting* | *Scent* | 28,341 | 8,090 |
| | *Longevity* | 29,532 | 6,899 |
| | *Sillage* | 29,393 | 7,038 |

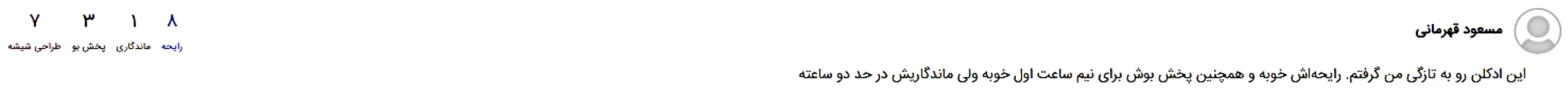

Fig. 6. Example of Sentiment Shift from Positive to Negative After User Vote Integration for both Longevity and Sillage (The user voted “1” for longevity and “3” for Sillage). The initial positive sentiment was influenced by the scent and initial sillage, but the user later expressed disappointment regarding its longevity. Translation of the comment: *"I recently got this perfume. Its fragrance is nice, and the projection is good for the first half hour, but its longevity is only about 2 hours."*

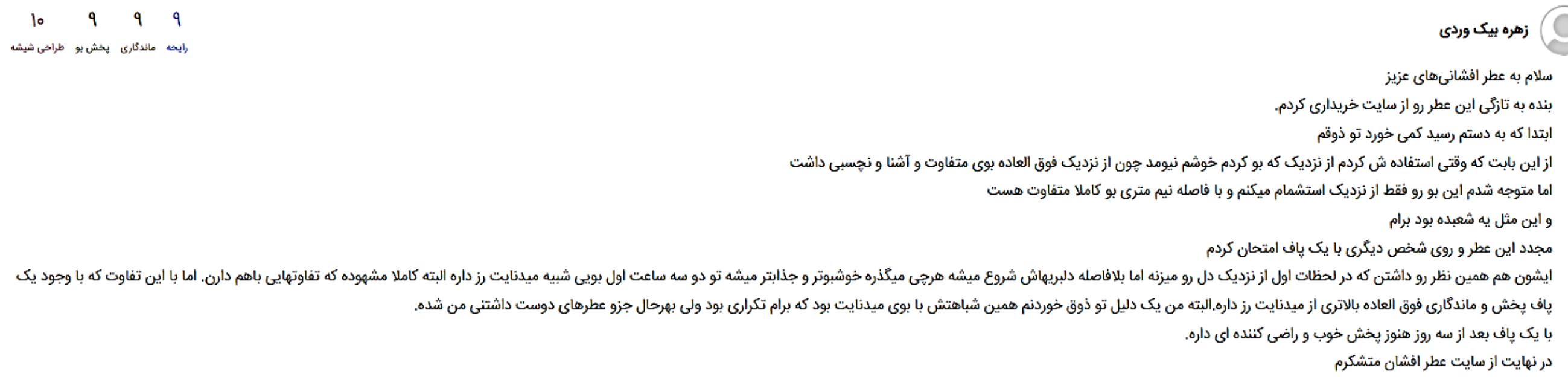

Fig. 7. Example of Sentiment Shift from Negative to Positive After User Vote Integration for All Categories.(The user voted “9” for all three essential categories.) The ParsBert model initially misclassified the comment due to its complexity and nuanced sentiment, as the user shared mixed emotions about the perfume's initial and later impressions. Translation of the comment: *"Hello, dear Atrafshan users. I recently purchased this perfume from the site. At first, I was a little disappointed because when I used it and smelled it up close, I didn't like it—it had a different, familiar, and unpleasant scent up close. But I realized that I only smell this scent when I am very close, and from half a meter away, the smell is completely different. It was like magic for me. I tried this perfume on someone else with one spray, and they had the same opinion—that it initially disappoints up close but then immediately starts to charm you. The more time passes, the more pleasant and attractive it becomes. In the first 2 to 3 hours, it has a scent somewhat like “Midnight Rose”, although it's clear that they have their differences. However, with just one spray, it has far better projection and longevity compared to “Midnight Rose”. Part of my initial disappointment was due to its similarity to “Midnight Rose”, which felt repetitive to me, but in any case, it has become one of my favorite perfumes. With just one spray, it still has good, satisfying projection after 3 days. Finally, I thank the Atrafshan website."*

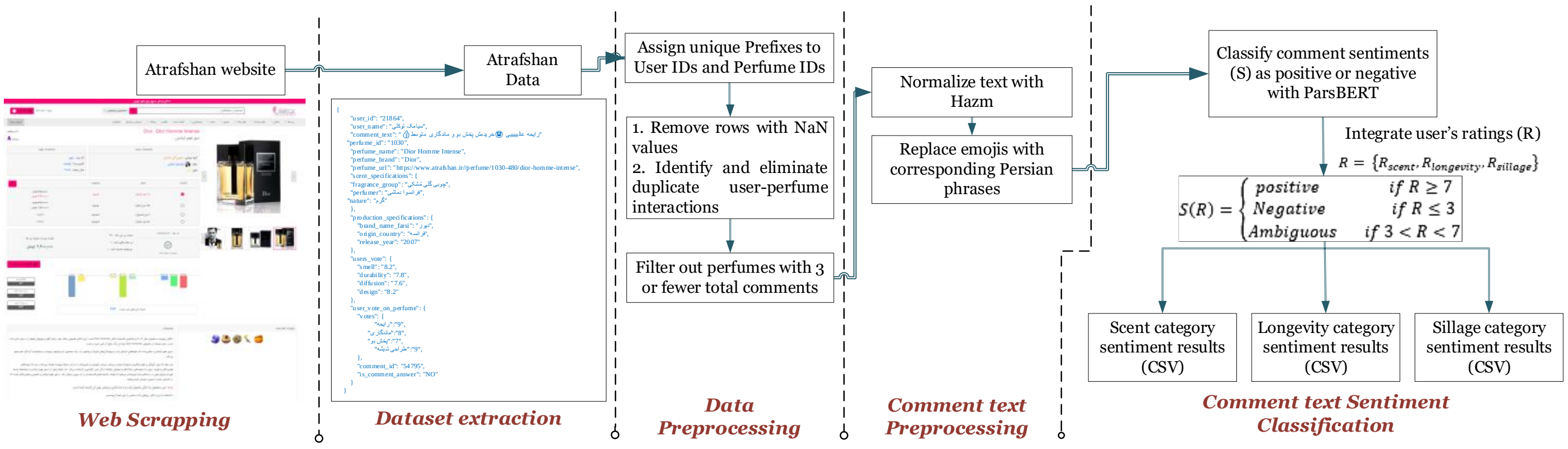

Fig. 8. Schematic Representation of Our Methodology: From Web Scraping to Comment Text Sentiment Classification.

## D. Network Construction

- **Bipartite User-Perfume Network:** The bipartite network is created by defining two separate sets of nodes that represent users and perfumes, where edges are formed solely based on positive comments left by users regarding specific perfumes. In this network, each user node is connected to perfume nodes to which they have provided favorable feedback, establishing a clear relationship that reflects user preferences. Initially, the graph was constructed using all comments, both positive and negative. After removing perfumes with three or fewer comments, the network comprised a total of 7,387 unique users and 1,104 unique perfumes, leading to 8,491 graph nodes and 35,910 edges.

After filtering for positive comments, we constructed three separate bipartite user-perfume networks based on sentiment analysis results for three key aspects of the perfume experience: Scent, Longevity, and Sillage. As mentioned in the sentiment analysis section, user votes were integrated into the sentiment analysis.

The final statistics for these networks are reported in Table IV. These refined bipartite networks provide a robust representation of user preferences across different dimensions of the perfume experience, laying the groundwork for further analysis, including the transformation into Perfume Co-Preference Networks for more advanced recommendations.

- **Perfume Co-Preference Network:** The perfume co-preference network is constructed using matrix multiplication to capture user preferences across different perfumes. This is done by taking the bipartite matrix, which links users to perfumes based on positive sentiment, and multiplying it by its transpose.

  Let the user-perfume bipartite matrix be represented as $A$, Where $n$ is the number of users, $m$ is the number of perfumes and each element $A_{ij}$ is 1 if user $i$ has commented positively on perfume $j$, and 0 otherwise.

$$A = \begin{bmatrix} a_{11} & a_{12} & \cdots & a_{1m} \\ a_{21} & a_{22} & \cdots & a_{2m} \\ \vdots & \vdots & \ddots & \vdots \\ a_{n1} & a_{n2} & \cdots & a_{nm} \end{bmatrix}$$

  First, the transpose of the user-perfume bipartite matrix, $A^T$, is calculated:

$$A^T = \begin{bmatrix} a_{11} & a_{21} & \cdots & a_{n1} \\ a_{12} & a_{22} & \cdots & a_{n2} \\ \vdots & \vdots & \ddots & \vdots \\ a_{1m} & a_{2m} & \cdots & a_{nm} \end{bmatrix}$$

  Now, $A^T$ has dimensions $m \times n$, where each row corresponds to a perfume and each column to a user. Next, matrix multiplication between $A^T$ and the original matrix $A$ is performed:

$$P = A^T \cdot A$$

  This matrix multiplication results in an $m \times m$ perfume-perfume adjacency matrix $P$, Where:

$$P_{jk} = \sum_{i=1}^{n} A_{ji}^T \cdot A_{ik}$$

  -$P_{jk}$ is the number of users who reviewed both perfume $j$ and perfume $k$.

  -$A_{ji}^T$ is the element in the $j$-th row and $i$-th column of $A^T$.

  -$A_{ik}$ is the element in the $i$-th row and $k$-th column of $A$.

Thus, the perfume co-preference matrix $P$ reflects the shared preferences among users, where the diagonal entries $P_{jj}$ represent the total number of users who commented positively on perfume $j$, and the off-diagonal entries $P_{jk}$ indicate co-preferences between perfume $j$ and perfume $k$.

Initially, the network contains self-loops, represented by the diagonal entries $P_{jj}$ of the perfume copreference matrix. These entries indicate the total number of positive comments each perfume has received. However, these self-loops are not useful for community detection, as they simply reflect a perfume's overall popularity rather than its relationship with other perfumes. To enhance the analysis, the self-loops ( $P_{jj}$) are removed, allowing a focus on the connections between different perfumes, which are represented by the off-diagonal entries $P_{jk}$.

After filtering for positive comments, three separate perfume co-preference networks were constructed from the three bipartite user-perfume networks, as described in the Bipartite User-Perfume Network section.

In Table IV., the statistics for these three perfume co-preference networks, derived from the respective bipartite graphs, are reported.

Next, the edges of the perfume co-preference graph were refined to better capture user sentiment by incorporating category-specific average ratings (such as Scent, Longevity, and Sillage) in addition to the original adjacency matrix values, which reflect the number of users who commented positively on both perfumes. For instance, an average Scent rating might be 8.2 based on 285 votes.

To determine the final edge weight $W_{jk}$ between two perfumes $j$ and $k$, we calculate it using the following formula:

$$W_{jk} = \left(0.6 \times R_j + 0.6 \times R_k\right) + \left(0.4 \times C_{jk}\right)$$

In this equation:

-$R_j$ and $R_k$ represent the category-specific average ratings (e.g., for Scent, Longevity, or Sillage) for perfumes $j$ and $k$ respectively.

-$C_{jk}$ represents the original adjacency value, indicating the number of users who commented positively on both perfumes $j$ and $k$.

This formula was developed through trial and error, experimenting with various formulas to achieve optimal outputs. Through this iterative process, we found that this formula provided the best results, both in terms of modularity score and in generating balanced detected communities. By weighting the edges in this manner, the strength of co-preference relationships is captured more effectively, improving the accuracy of community detection.

Notably, the edge weight refinement technique contributed to a significant increase in the modularity score across all of our community detection techniques compared to the primary network. This refinement allows for the identification of meaningful clusters of perfumes that are often preferred together, offering deeper insights into consumer preferences within the fragrance industry.

TABLE IV. STATISTICS OF THE BIPARTITE USER-PERFUME NETWORKS AND CORRESPONDING PERFUME CO-PREFERENCE NETWORKS

| | | | *Nodes* | | *Edges* | *Weighted* | *Undirected* |
|---|---|---|---|---|---|---|---|
| | | | **User** | **Perfume** | | | |
| *User-Perfume Graph* | *Primary* | | 7,387 | 1,369 | 36,434 | NO | YES |
| | *Removing perfumes With three or fewer comments* | | 7,387 | 1,104 | 35,910 | NO | YES |
| | *Positive Comments only* | *Scent* | 6,191 | 1,008 | 27,402 | NO | YES |
| | | *Longevity* | 6,493 | 1,015 | 28,600 | NO | YES |
| | | *Sillage* | 6,477 | 1,012 | 28,458 | NO | YES |
| *Perfume Co-Preference (Without self-loop)* | | *Scent* | - | 1,008 | 169,916 | YES | YES |
| | | *Longevity* | - | 1,015 | 172,884 | YES | YES |
| | | *Sillage* | - | 1,012 | 173,206 | YES | YES |

### E. *Community Detection*

In this study, various community detection algorithms were applied across the three Perfume Co-Preference Networks (scent, longevity, and sillage), covering multiple approaches to ensure comprehensive clustering based on user preferences.

For modularity-based methods, we employed both the Louvain and Fast Greedy algorithms. From the category of dynamic algorithms, the Walktrap and Spinglass algorithms were utilized. Additionally, spectral clustering was applied as a spectral-based method for detecting communities.

Modularity was calculated for all community detection algorithms to assess the quality of the partitions generated, ensuring that the grouping of perfumes was meaningful based on user preferences.

- **Weighting Scheme and Network Refinement:** In all methods, the same weighting scheme was used in each co-preference network, following the 60-40 rule (as outlined in the Perfume Co-Preference Network construction). Several filtering rules were applied consistently across all networks to clean and refine the graph before community detection:
  - **Removal of low-weight edges:** Edges with weights less than or equal to 3 were removed to minimize noise.
  - **Elimination of isolated nodes:** Perfumes with no connections (i.e., perfumes with a degree of zero) were removed to prevent skewing the results.

- **Edge Weight Normalization:** To ensure that the edge weights reflected the relative importance of perfume connections accurately, the weights were further normalized based on the degrees of the nodes (i.e., perfumes) at each end of the edge. The formula for normalization was:

$$w_{\text{normalized}} = \frac{w_{\text{combined}}}{\text{degree}_u \times \text{degree}_v}$$

where $\text{degree}_u$ and $\text{degree}_v$ represent the degrees of the two connected perfumes. This step ensured that highly connected nodes (popular perfumes) did not disproportionately influence the clustering results, while still maintaining the integrity of the co-preference relationships.

- **Implementation and Output of Community Detection Algorithms:** After constructing and filtering the networks, the aforementioned community detection algorithms were applied. The results, including the number of detected communities, were saved in the following formats:
  - **PNG files**: Visual plots of the detected communities.
  - **CSV files**: Community detection outcomes, including the associated perfume names and their respective communities.
  - **GraphML files**: Containing both the graph structure and the embedded community information within the nodes.

## IV. RESULTS

This section provides a comprehensive overview of the experimental approach and outcomes, focusing on the application of five community detection techniques: Louvain (with a resolution of 2), Fast Greedy, Walktrap, Spinglass, and Spectral Clustering. These techniques were implemented across seven networks.

The first network was the Primary network, which included both positive and negative comments, with perfumes that had three or fewer comments being removed.

The next set of networks integrated user's Comment Sentiment Analysis, using only positive comments, and was applied individually to the attributes of Scent, Longevity, and Sillage.

Finally, we applied both the user's Comment Sentiment Analysis, using only positive comments, and refined the edge weights with average total user ratings to create three additional networks, again focused on the attributes of Scent, Longevity, and Sillage.

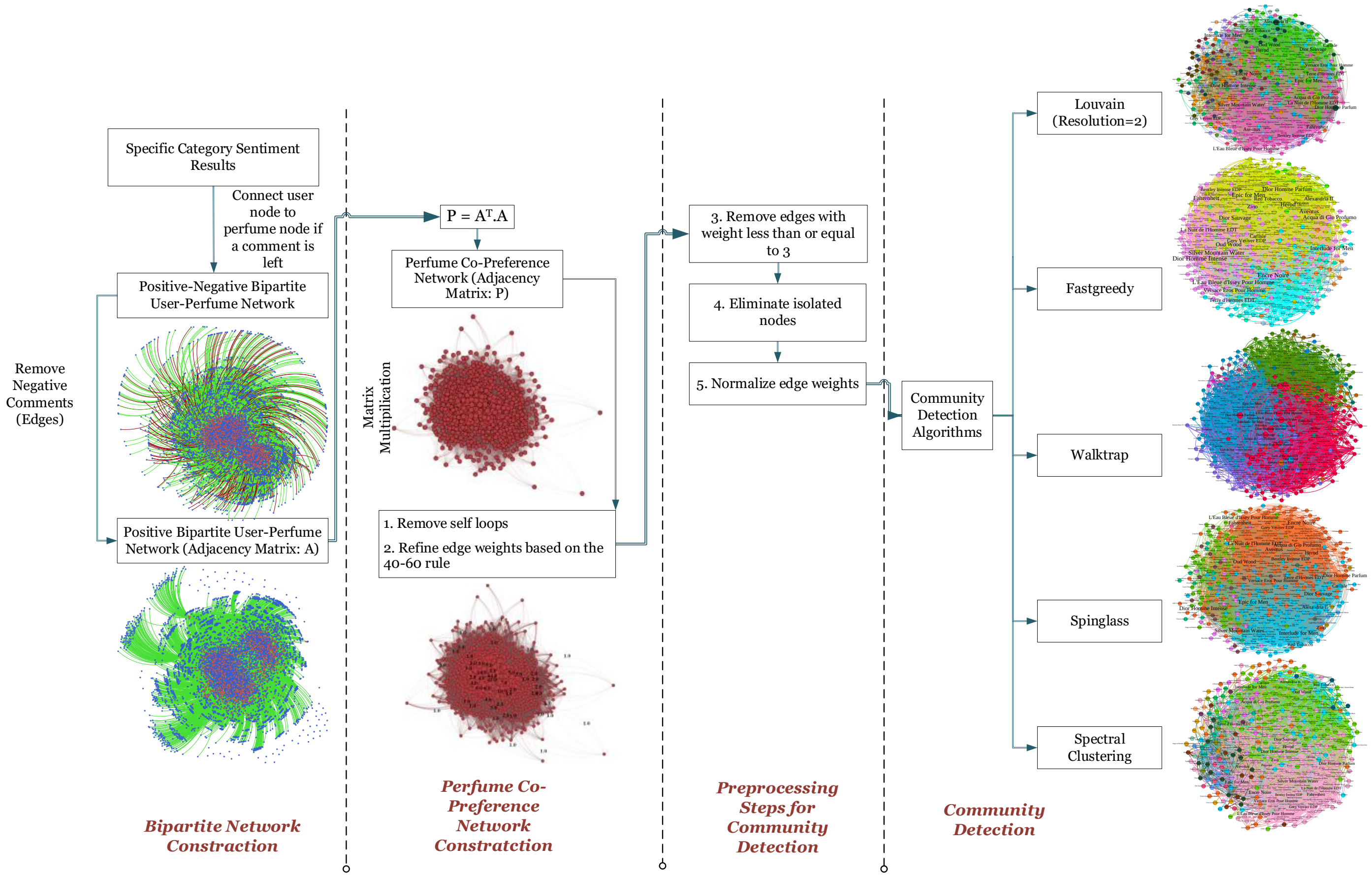

Fig. 9. Schematic Representation of Our Methodology: From Network Construction to Community Detection. Note that this process is conducted individually for the sentiment results specific to each category.

- **Modularity Score Improvements Impact of Sentiment Analysis and Edge Weight Refinement:**
From the results in Table V., we observed a clear trend in modularity score improvement. For example, the modularity score for the Louvain method in the Primary network was 0.41. When the users' Comment Sentiment Analysis was incorporated, the modularity scores for the Louvain method increased to between 0.44 and 0.46, indicating an approximate percentage increase of 7.3% to 12.2% compared to the Primary network. This change suggests that the community detection algorithms were better able to identify meaningful clusters of perfumes.

This improvement was even more significant when the average total rating of perfumes in edge weights was added, resulting in further increases in modularity, with the Louvain method achieving scores between 0.52 and 0.53. This represents an approximate percentage increase of 13% to 20.5% from the previous scores with Comment Sentiment Analysis. Overall, the increase in modularity from the Primary network score of 0.41 to the final scores of 0.52 to 0.53 indicates an approximate percentage increase of 27% to 29.3%.

This progression produced more well-defined community structures. Similar trends were observed across the other community detection methods, including Fast Greedy, Walktrap, Spinglass and Spectral Clustering.

For the three networks that included both the users' Comment Sentiment Analysis, using only positive comments, and refined edge weights with average total user ratings, the modularity score averaged 0.53 across all community detection techniques. The lowest modularity was 0.48, belonging to the Walktrap method, while the highest modularity was 0.58, achieved by Spinglass. This modularity score indicates that the integration of sentiment analysis and refined edge weights contributed to the formation of more cohesive and well-defined communities, effectively capturing user preferences within the perfume co-preference networks.

In comparison, the average modularity score for the primary network across all methods was 0.43, resulting in an approximate percentage increase of 23.3% from the primary network to the networks incorporating sentiment analysis and refined edge weights.

These findings highlight that by implementing sentiment analysis on user comments, constructing the network with positive comments only, and refining edge weights with average total user ratings, the performance of community detection algorithms was significantly enhanced, leading to clearer partitions and stronger associations within the perfume co-preference networks.

- **Performance of Community Detection Techniques:** Among the community detection techniques, Spinglass, Fast Greedy, and Louvain achieved the best modularity scores, with Spinglass recording the highest modularity at 0.58, Fast Greedy at 0.56, and Louvain at 0.53.

  In contrast, Walktrap and Spectral Clustering performed more poorly compared to the other methods. Walktrap recorded a modularity score of 0.48 to 0.49, while Spectral Clustering's performance ranged from 0.49 to 0.51.

  This reduced performance could be attributed to their sensitivity to network structure. Additionally, the community structures identified by these methods may not have aligned as closely with user preferences as those detected by the other techniques, further contributing to their lower scores.

## V. CONCLUSIONS

In this study, we have successfully constructed a perfume co-preference network utilizing user reviews from the Persian retail platform Atrafshan.

By applying community detection algorithms to this network, we uncovered valuable clusters of similar perfumes, providing significant insights into user sentiment and preferences within the fragrance industry.

Our innovative approach to sentiment analysis, which integrated emojis and a user voting system, allowed for a more nuanced understanding of user feedback.

Furthermore, refining edge weights based on average total user ratings for attributes such as Scent, Longevity, and Sillage enhanced the accuracy and meaningfulness of our network, leading to improved insights into consumer behavior.

Notably, our inclusion of sentiment classification—constructing the network based solely on positive comments—combined with the refinement of edge weights, contributed to a significant increase in the modularity score across all community detection techniques compared to the primary network.

These enhancements facilitated clearer and more meaningful partitions within the network, resulting in stronger associations among perfumes within the identified clusters based on shared user preferences. This approach not only improves the precision of perfume recommendations but also provides deeper insights into consumer preferences and behaviors in the fragrance industry.

TABLE V. THE STATISTICS OF COMMUNITY DETECTIONS

| *Method* | *Network Construction Type* | | *Modularity* | *Number of Communities Detected* |
|---|---|---|---|---|
| *LOUVAIN (Resolution=2)* | *Primary* | | 0.41 | 30 |
| | *Integration of User's comment Sentiment Analysis* | *Scent* | 0.45 | 30 |
| | | *Longevity* | 0.46 | 34 |
| | | *Sillage* | 0.44 | 34 |
| | *Integration of User's comment Sentiment Analysis + Perfume Total Voting in edge weights* | *Scent* | 0.53 | 21 |
| | | *Longevity* | 0.52 | 24 |
| | | *Sillage* | 0.52 | 24 |
| *FAST GREEDY* | *Primary* | | 0.48 | 14 |
| | *Integration of User's comment Sentiment Analysis* | *Scent* | 0.51 | 18 |
| | | *Longevity* | 0.52 | 16 |
| | | *Sillage* | 0.51 | 15 |
| | *Integration of User's comment Sentiment Analysis + Perfume Total Voting in edge weights* | *Scent* | 0.54 | 21 |
| | | *Longevity* | 0.55 | 19 |
| | | *Sillage* | 0.56 | 17 |
| *WALK TRAP* | *Primary* | | 0.44 | 3 |
| | *Integration of User's comment Sentiment Analysis* | *Scent* | 0.46 | 3 |
| | | *Longevity* | 0.47 | 3 |
| | | *Sillage* | 0.47 | 3 |
| | *Integration of User's comment Sentiment Analysis + Perfume Total Voting in edge weights* | *Scent* | 0.49 | 4 |
| | | *Longevity* | 0.48 | 4 |
| | | *Sillage* | 0.48 | 3 |
| *SPIN GLASS* | *Primary* | | 0.49 | 8 |
| | *Integration of User's comment Sentiment Analysis* | *Scent* | 0.52 | 12 |
| | | *Longevity* | 0.57 | 10 |
| | | *Sillage* | 0.52 | 12 |
| | *Integration of User's comment Sentiment Analysis + Perfume Total Voting in edge weights* | *Scent* | 0.58 | 10 |
| | | *Longevity* | 0.58 | 11 |
| | | *Sillage* | 0.58 | 10 |
| *SPECTRAL CLUSTERING* | *Primary* | | 0.31 | 29 |
| | *Integration of User's comment Sentiment Analysis* | *Scent* | 0.43 | 29 |
| | | *Longevity* | 0.40 | 27 |
| | | *Sillage* | 0.40 | 27 |
| | *Integration of User's comment Sentiment Analysis + Perfume Total Voting in edge weights* | *Scent* | 0.49 | 18 |
| | | *Longevity* | 0.51 | 19 |
| | | *Sillage* | 0.50 | 19 |

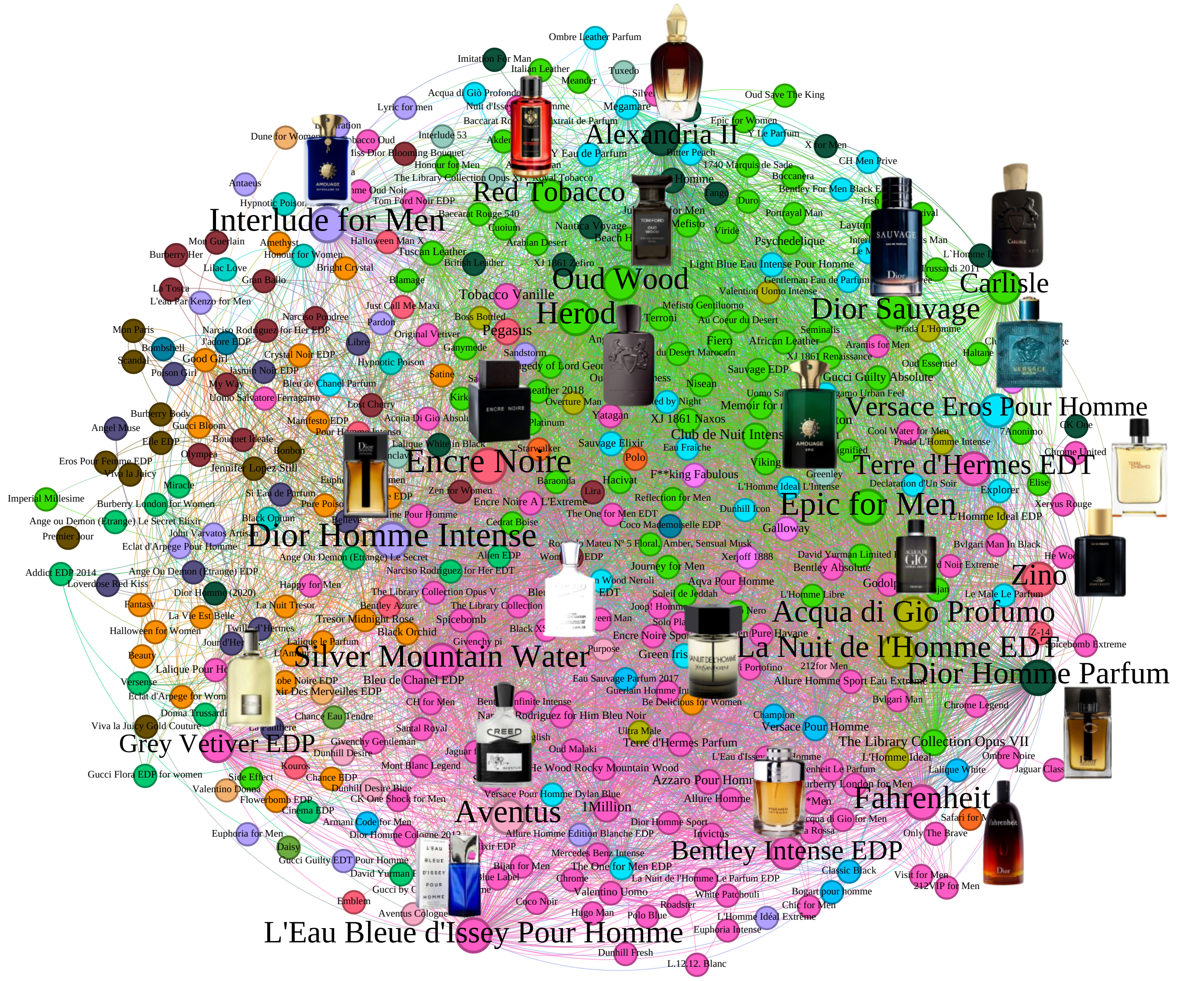

Fig. 10. Visualization of Communities Detected by the Louvain Method in the Scent Network, Incorporating User Comment Sentiment Analysis and Perfume Total Voting in Edge Weights.

TABLE VI. SUMMARY OF PERFUME COMMUNITIES DETECTED BY THE LOUVAIN METHOD IN THE SCENT NETWORK, INCORPORATING USER COMMENT SENTIMENT ANALYSIS AND PERFUME TOTAL VOTING IN EDGE WEIGHTS.

| *Community* | *Population(perfumes)* | *Members* |
|---|---|---|
| C0 | 14 | 'Gran Ballo', 'Lira', 'La Tosca', 'Olympea', 'Bouquet Ideale', 'Narciso Rodriguez for Her EDP', 'Lost Cherry', 'Zen for Women', 'Narciso Poudree', 'Miss Dior Blooming Bouquet', 'Womanity EDP', 'My Way', 'Mon Guerlain', 'Burberry Her' |
| C1 | 2 | 'Chance Eau Tendre', 'Daisy' |
| C2 | 6 | 'Purpose', 'Dunhill Desire', 'Aventus', '212for Men', 'Dunhill Desire Blue', 'Aventus Cologne' |
| C3 | 27 | 'Le Male Le Parfum', 'Bvlgari Man Wood Neroli', 'Versace Pour Homme Dylan Blue', 'L'Homme Ideal L'Intense', 'Eau Fraiche', 'Bitter Peach', 'Dunhill Icon', 'Wanted by Night', 'Uomo Salvatore Ferragamo Urban Feel', 'Versace Eros Pour Homme', 'The One for Men EDP', 'Le Male', 'Bleu de Chanel Parfum', 'Megamare', 'Light Blue Eau Intense Pour Homme', 'Explorer', 'Declaration d'Un Soir', 'Acqua di Giò Profondo', 'Green Irish Tweed', 'Gentleman Eau de Parfum Reserve Privée', 'Sauvage Elixir', 'CH Men Prive', 'Ombre Leather Parfum', 'Bentley For Men Black Edition', 'Y Eau de Parfum', 'Bleu de Chanel EDT', 'Y Le Parfum' |
| C4 | 7 | 'Enclave', 'Greenley', 'Jubilation for Men', 'The Library Collection Opus XIV Royal Tobacco', 'Interlude 53', 'Search', 'Tuxedo' |
| C5 | 74 | 'Viride', 'Dior Sauvage', 'Cuoium', 'Gucci Guilty Absolute', 'Journey for Men', 'Uomo Trussardi 2011', '02 L'Air du Desert Marocain', 'Arabian Desert', 'Beach Hut Man', 'Carlisle', 'Psychedelique', 'Mefisto Gentiluomo', 'Rosendo Mateu Nº 5 Floral, Amber, Sensual Musk', 'Portrayal Man', 'Blamage', 'David Yurman Limited Edition', 'Au Coeur du Desert', 'Oud for Greatness', 'Nisean', 'Side Effect', 'Soleil de Jeddah', 'Reflection for Men', 'Club de Nuit Intense for men', 'The Tragedy of Lord George', 'Angels' Share', 'Honour for Men', 'Fiero', 'Imperial Millesime', 'Laudano Nero', 'Cedrat Boise', 'L'Homme Libre', 'Ambre Sultan', 'Epic for Men', 'Kirke', 'Boccanera', 'Akdeniz', 'Terroni', 'Meander', 'African Leather', 'Seminalis', 'Oajan', 'Italian Leather', 'Ganymede', 'Red Tobacco', 'Oud Essentiel', 'Hacivat', 'Memoir for men', 'XJ 1861 Renaissance', 'Percival', 'Tuscan Leather', 'Elise', 'Baraonda', 'Oud Wood', 'Sauvage EDP', 'Baccarat Rouge 540', 'Layton Exclusif', 'XJ 1861 Naxos', 'Egoiste Platinum', 'Dignified', 'Haltane', 'Mefisto', 'Interlude Black Iris Man', 'Layton', |
| | | '7Anonimo', 'The Library Collection Opus VII', '1740 Marquis de Sade', 'Ombre Leather 2018', 'Epic for Women', 'Duro', 'Viking', 'Irish Leather', 'Baccarat Rouge 540 Extrait de Parfum', 'Herod', 'Oud Save The King' |
| C6 | 94 | 'Santal Royal', 'CH for Men', 'Bentley Infinite Intense', 'Jaguar for Men', 'Allure Homme Sport', 'Black XS for Men', 'Yatagan', 'The Library Collection Opus VIII', 'Coco Noir', 'Neroli Portofino', 'Bijan for Men', 'Guerlain Homme Intense EDP', 'Tobacco Vanille', 'Invictus', 'Polo Blue', 'Tom Ford Noir EDP', 'Original Vetiver', 'Silver Mountain Water', '1Million', 'Allure Homme Sport Eau Extreme', 'Oud Malaki', ‘La Nuit de l'Homme Le Parfum EDP’, 'Halloween Man', 'A*Men', 'CK One Shock for Men', 'Lalique Pour Homme EDP', 'Hugo Man', 'Mine Pour Homme', 'Givenchy pi', 'Burberry London for Men', 'Mercedes Benz Intense', 'The Library Collection Opus V', 'Acqua Di Gio Absolu', 'Solo Platinum', 'Only The Brave', 'Shagya', 'Bentley Absolute', 'Encre Noire Sport', 'Fahrenheit', 'Jaguar Classic Gold', 'Azzaro Pour Homme', 'Uomo Salvatore Ferragamo', 'Cool Water for Men', 'He Wood Rocky Mountain Wood', 'Happy for Men', 'White Patchouli', ‘L'Eau d'Issey Pour Homme’, 'Aromatics Elixir EDP', 'Bentley Intense EDP', 'Spicebomb', 'Fahrenheit Le Parfum', 'Narciso Rodriguez for Him Bleu Noir', 'Silver Scent', 'Chrome', ‘L'Eau Bleue d'Issey Pour Homme’, 'Pour Homme Oud Noir', 'Aqva Pour Homme', 'Aramis for Men', ‘Terre d'Hermes EDT’, 'Tom Ford Noir Extreme', 'Blue Label', 'The One for Men EDT', 'Valentino Uomo', 'Dunhill Fresh', 'Dior Homme Cologne 2013', 'Joop! Homme', 'L.12.12. Blanc', 'A*Men Pure Havane', ‘Encre Noire A L'Extreme’, 'Grey Vetiver EDP', ‘Terre d'Hermes Parfum’, 'Bvlgari Man', 'Tobacco Oud', 'Bvlgari Man In Black', 'Euphoria Intense', 'Visit for Men', 'Spicebomb Extreme', 'Gucci by Gucci Pour Homme', 'Roadster', 'Bleu de Chanel EDP', ‘Nuit d'Issey Pour Homme’, 'Chrome Legend', 'Acqua di Gio for Men', 'Xeryus Rouge', '212VIP for Men', 'Mont Blanc Legend', 'Chrome United', 'He Wood', 'Chic for Men', 'Ombre Noire', 'Luna Rossa', 'Givenchy Gentleman', 'Godolphin', 'Dior Homme Sport' |
| C7 | 12 | 'Miracle', 'Alien EDP', 'Narciso Rodriguez for Her EDT', 'Ange ou Demon (Etrange) Le Secret Elixir', 'David Yurman Fragrance', 'Gucci Flora EDP for women', 'Cinema EDP', 'Ange Ou Demon (Etrange) Le Secret', 'Burberry London for Women', 'Donna Trussardi 2011', 'Versense', 'Addict EDP 2014' |
| C8 | 11 | 'Si Eau de Parfum', 'Libre', 'La Panthere', ‘Jour d'Hermes’, 'Ange Ou Demon (Etrange) EDP', 'Loverdose Red Kiss', 'Twilly d’Hermes', 'Jasmin Noir EDP', 'Poison Girl', 'Believe', 'Angel Muse' |
| C9 | 11 | 'L'Homme Ideal EDP', 'Ultra Male', 'Dirty English', 'Overture Man', ‘Prada L'Homme’, ‘La Nuit de l'Homme EDT’, 'Valentino Uomo Intense', ‘L'Homme Ideal’, 'Boss Bottled', ‘Prada L'Homme Intense’, 'Acqua di Gio Profumo' |
| C10 | 15 | 'John Varvatos Artisan', 'Erba Pura', 'Interlude for Men', 'Pardon', 'Allure Homme Edition Blanche EDP', ‘Eclat d'Arpege Pour Homme’, ‘L'eau Par Kenzo for Men’, 'Sandstorm', 'Lyric for men', 'Dior Homme Intense', 'Declaration', 'Euphoria for Men', 'Antaeus', ‘L'Homme Idéal Extrême’, 'Gucci Guilty EDT Pour Homme' |
| C11 | 3 | 'Starwalker', 'Polo', 'Safari for Men' |
| C12 | 24 | 'Be Delicious for Women', ‘Eclat d'Arpege for Women’, 'Bright Crystal', 'Gucci Bloom', 'Chance EDP', 'Euphoria for Women', 'Lalique le Parfum', 'Idylle EDP', 'Halloween for Women', 'Manifesto EDP', 'Fantasy', 'La Petite Robe Noire EDP', 'Amethyst', 'Flowerbomb EDP', 'Beauty', 'La Vie Est Belle', 'Satine', 'La Nuit Tresor', 'Black Orchid', 'Pure Poison', 'L'Amour', 'Crystal Noir EDP', 'Good Girl', 'Tresor Midnight Rose' |
| C13 | 16 | 'Dior Homme Parfum', 'XJ 1861 Zefiro', 'Lalique White in Black', 'British Leather', 'Alexandria II', 'Nautica Voyage', 'CK One', 'L`Homme Lacoste', 'X for Men', 'Pour Un Homme', 'Tango', 'Dior Homme (2020)', 'Club De Nuit Sillage', 'H24', 'Kalan', 'Imitation For Man' |
| C14 | 10 | 'Zino', 'Bentley Azure', 'Just Call Me Maxi', 'Pour Homme Intenso', 'Kouros', 'Halloween Man X', 'Encre Noire', 'Z-14', 'Pegasus', 'Emblem' |
| C15 | 5 | 'Hypnotic Poison', 'Hypnotic Poison EDP', 'Lilac Love', 'Black Opium', 'Honour for Women' |
| C16 | 5 | 'Sadonaso', 'Xerjoff 1888', 'Galloway', 'Ispazon', 'F**king Fabulous' |
| C17 | 3 | 'Coco Mademoiselle EDP', 'Bombshell', ‘J'adore EDP’ |
| C18 | 3 | 'Elixir Des Merveilles EDP', 'Valentino Donna', 'Dune for Women' |
| C19 | 10 | 'Viva la Juicy', 'Scandal', 'Eros Pour Femme EDP', 'Elle EDP', 'Burberry Body', 'Premier Jour', 'Bonbon', 'Jennifer Lopez Still', 'Mon Paris', 'Viva la Juicy Gold Couture' |
| C20 | 7 | 'Classic Black', 'Lalique White', 'Champion', 'Armani Code for Men', 'Eau Sauvage Parfum 2017', 'Versace Pour Homme', 'Bogart pour homme' |

The findings from this research underscore the importance of community detection in the analysis of perfume networks, paving the way for more personalized recommendation systems and targeted marketing strategies. By identifying clusters of similar perfumes, businesses can better align their product offerings with consumer preferences, thereby enhancing customer satisfaction and loyalty. Looking ahead, several avenues for future work are promising. A perfume recommender system could be implemented based on user profiles derived from the co-preference network, offering personalized recommendations tailored to individual preferences and enhancing the overall user experience. Moreover, future research could explore the application of the perfume co-preference network in real-time recommendation systems, leveraging dynamic user interactions and feedback to continually refine and adapt recommendations. Additionally, developing a model specifically designed for sentiment analysis on perfume comments and reviews could yield more accurate sentiment classifications, ultimately enriching the insights gained from user feedback.

Lastly, investigating the influence of external factors, such as marketing campaigns or seasonal trends, on perfume preferences could provide valuable insights into the ever-evolving landscape of consumer behavior in the fragrance market.

Overall, this research represents a significant step forward in understanding consumer preferences in the fragrance industry, and we are optimistic about the potential impact of future work in this area.

## VI. Acknowledgments

The authors would like to express their heartfelt gratitude to Ms. Sajedeh Olyaee for her collaboration on an earlier project.